\documentclass{aa}  

\usepackage{graphicx}
\usepackage{txfonts}

\newcommand{\hi}{\ensuremath{\mathrm{H\,\textsc{i}}}}
\begin{document}

   \title{The satellite galaxy plane of NGC\,4490 in light of $\Lambda$CDM}

   \subtitle{Sparsity of similarly extreme analogs and a possible role of satellite pairs}

   \author{Marcel S. Pawlowski
          \inst{1}
          \and
          Oliver Müller \inst{2}
          \and
          Salvatore Taibi \inst{1}
          \and
          Mariana P. Júlio \inst{1, 3}
          \and
          Kosuke Jamie Kanehisa \inst{1, 3}
          \and \\
          Nick Heesters \inst{2}
          }

   \institute{Leibniz-Institute for Astrophysics Potsdam (AIP), An der Sternwarte 16, 14482 Potsdam, Germany\\
              \email{mpawlowski@aip.de}
         \and
             Institute of Physics, Laboratory of Astrophysics, Ecole Polytechnique Fédérale de Lausanne (EPFL), 1290 Sauverny, Switzerland
        \and
             Institut für Physik und Astronomie, Universität Potsdam, Karl-Liebknecht-Straße 24/25, D-14476 Potsdam, Germany
             }

   \date{Received May 6, 2024; accepted May 6, 2024}

 
  \abstract
   {The system of galaxies around NGC\,4490 was recently highlighted to display a flattened, kinematically correlated structure reminiscent of planes of satellite galaxies around other hosts.}
   {Since known satellite planes are in tension with expectations from cosmological simulations in the Lambda Cold Dark Matter ($\Lambda$CDM) model, we aim to quantitatively assess for the first time the tension posed by the NGC\,4490 system.}
   {We measured the on-sky flattening as the major-to-minor axis ratio $b/a$\ of the satellite distribution and their line-of-sight kinematic correlation. Analogs to the system were selected in the IllustrisTNG-50 simulation and their flattening and correlation were similarly measured.}
   {We confirm the strong kinematic coherence of all 12 observed satellite objects with available line-of-sight velocities (of 14 in total): the northern ones approach and the southern ones recede relative to the host. The spatial distribution of all 14 objects is substantially flattened with $b/a = 0.38$\ (0.26 considering only the 12 objects with available velocities). Such extreme arrangements are rare in the $\Lambda$CDM simulation, at a level of 0.21 to 0.35\%. This fraction of analogs would drop further if at least one of the two satellite objects without velocities is confirmed to follow the kinematic trend, and would become zero if both are rejected as non-members. We also identify a likely galaxy pair in the observed system, and find a similar pair in the best-matching simulated analog.}
   {Our measurements establish NGC\,4490 as another strong example of a satellite plane in the Local Volume. This emphasizes that planes of satellites are a more general issue faced by $\Lambda$CDM also beyond the Local Group. The tension with typical systems drawn from simulations suggests that the observed one requires a specific formation scenario, potentially connected to the larger-scale galaxy alignment in its vicinity.
The presence of galaxy pairs in the observed and a simulated system hints at the importance such groupings may have to understand satellite planes.}


   \maketitle
%

\section{Introduction}
 
Already \citet{1976MNRAS.174..695L} found that the satellite galaxies (and some other objects) in the halo of the Milky Way distribute along a common great circle, akin to lying in a common plane perpendicular to the Galaxy. The discovery of additional, fainter satellite galaxies further supported this peculiar Vast Polar Structure (VPOS, \citealt{2012MNRAS.423.1109P, 2016MNRAS.456..448P}). Proper motion measurements indicate that the members of the VPOS co-orbit \citep{2021ApJ...916....8L, 2024A&A...681A..73T}. A similar, highly flattened structure consisting of about half of the satellites of the Andromeda Galaxy (M31) was identified by \citet{2013Natur.493...62I}. The edge-on orientation of this satellite plane allowed them to conclude from a strong line-of-sight velocity coherence -- on-plane satellites on one side of M31 predominantly approach us while those on the opposite side recede relative to the host -- that members of this structure potentially co-orbit, similar to the satellites in the VPOS. First proper motion measurements indicate that indeed at least two of the observed satellites orbit along the spatially flattened structure in a common direction \citep{2020ApJ...901...43S}, strengthening the picture of a co-rotating plane of satellites.
Outside of the Local Group, the most prominent and well-studied flattened satellite structure today is the one hosted by Centaurus\,A. It was first reported as a spatially flattened arrangement of satellite galaxies in approximately edge-on orientation \citep{2015ApJ...802L..25T}, and then confirmed to also display line-of-sight velocity coherence indicative of possible rotation along this spatial plane \citep{2018Sci...359..534M,2021A&A...645L...5M,2022A&A...662A..57M,2023MNRAS.519.6184K}.
 
   \begin{figure*}
   \sidecaption
   \includegraphics[width=12cm]{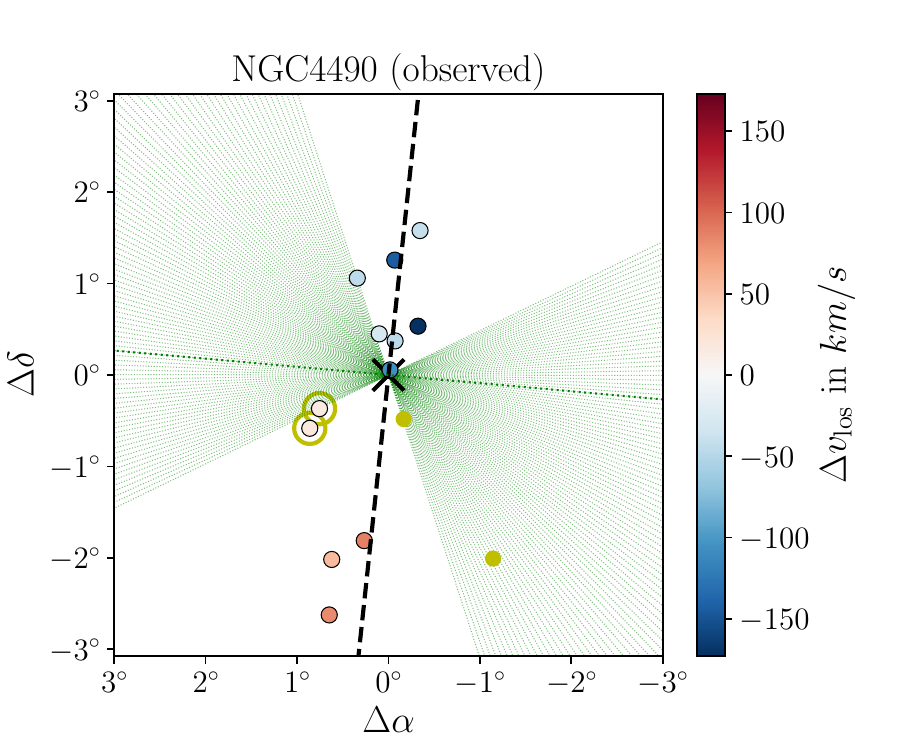}
      \caption{On-sky distribution of the NGC\,4490 system in equatorial coordinates centered on the host. Objects are color-coded by their velocity relative to that of NGC\,4490, and the two objects without available velocity measurements are shown in yellow. The major axis of the projected distribution is indicated with a dashed black line. The green dotted lines (in $1^\circ$\ increments) indicate the separation lines maximizing the kinematic coherence (all northern objects more blue-shifted, all southern objects more red-shifted than the host). This split maximizing the kinematic coherence aligns well with the distribution's minor axis (thicker green dotted line), as expected for a rotating, flattened structure, and similar to the M31 and Centaurus\,A satellite planes. The likely pair of dwarf galaxies is highlighted by yellow circles.
              }
         \label{FigOnSky}
   \end{figure*} 
 
These observed planes of satellite galaxies were identified as a challenge to the $\Lambda$CDM model of cosmology, a debate initiated by \citet{2005A&A...431..517K} and raging in the literature since \citep[e.g.,][]{2012PASA...29..395K, 2014MNRAS.442.2362P, 2015MNRAS.452.3838C, 2021NatAs...5.1188B, 2021NatAs...5.1185P}. Many studies compare the observed systems with analogs in cosmological simulations to determine the expected frequency of such features, and find them to be very rare. A majority of publications thus far focus on the Milky Way system \citep{2014MNRAS.442.2362P, 2018MNRAS.476.1796S, 2020MNRAS.491.3042P, 2021MNRAS.504.1379S, 2023MNRAS.520.3937P, 2023NatAs...7..481S, 2023ApJ...954..128X}, but results for M31 \citep{2014ApJ...784L...6I, 2021ApJ...923...42P} and CenA \citep{2018Sci...359..534M,2021A&A...645L...5M} are similar: analogs as extreme as the observed systems, employing metrics to measure both flattening and kinematic coherence, typically occur in less than one in several hundred simulated systems. This mismatch has become known as the planes of satellite galaxies problem of cosmology (for a review, see \citealt{2018MPLA...3330004P}).

Yet, these are only three out of countless satellite systems in the Universe, and for each of the three specific formation scenarios were proposed \citep[e.g.,][]{2013MNRAS.431.3543H, 2015ApJ...809...49B, 2016ApJ...818...11S, 2018A&A...614A..59B, 2020MNRAS.498.2766W, 2021ApJ...923..140G, 2024MNRAS.527..437V, 2024arXiv240108143S} or critisized \citep[e.g.,][]{2009ApJ...697..269M, 2019ApJ...875..105P, 2022ApJ...932...70P, 2023MNRAS.524..952K}. This raises the question whether these could simply be exceptional, special systems and that we are cosmically lucky to reside in one of them. To address this -- admittedly unsatisfactory but nevertheless not infrequently voiced -- possibility, it is worth investigating whether similar structures can be found around more distant hosts. Satellite-plane-like structures have now been proposed for a number of other hosts, including M\,81 \citep{2013AJ....146..126C,2024arXiv240116002M}, M\,101 \citep{2017A&A...602A.119M}, NGC\,253 \citep{2021A&A...652A..48M}, or NGC\,2750 \citep{2021ApJ...917L..18P}. Additionally, there are indications that such structures are rather frequent in larger statistical samples of systems \citep{2021A&A...654A.161H,2014Natur.511..563I, 2024NatAs.tmp...26G}.
However, most of these cases have not been studied in depth, or were not shown to be significantly offset from cosmological expectations. The latter is often hampered by a lack of observational constraints such as velocity information for a majority of dwarf satellites.
 
Recently, \citet{2024MNRAS.528.2805K} presented the NGC\,4490 galaxy and its surrounding objects as another example of a flattened and kinematically coherent structure similar to the known planes of satellite galaxies. They show that the on-sky distribution of the objects associated with the NGC\,4490 group is flattened along the Supergalactic plane, and exhibits a strong velocity coherence consistent with rotation that is also traced by the extended \hi\ structure surrounding the host. They furthermore report the discovery of an additional stellar system, a diffuse Plume, that is aligned with the flattened structure. However, no quantitative information on the flattening of the system or its significance has been reported, nor have any analogs in simulations been searched or any comparison has been made with cosmological expectations for the flattening and kinematic coherence of the system. This places the suggestion of \citet{2024MNRAS.528.2805K} that the NGC\,4490 system supports the notion that the planes of satellite galaxies problem is a major challenge for $\Lambda$CDM on somewhat unsure footing. In the following, we set out to rectify this by comparing the observed system with analogs selected from a cosmological simulation, to determine the frequency of similarly extreme systems. In doing so we follow a methodology analogous to previous studies, such as on Centaurus\,A \citep{2021A&A...645L...5M}. To further maximize comparability with the previously studied cases of planes of satellite galaxies, we also opt to base this comparison on the same simulation suite, specifically the IllustrisTNG simulations \citep{2019ComAC...6....2N}.


\section{The observed NGC\,4490 system}

   \begin{figure}
   \centering
   \includegraphics[width=\hsize]{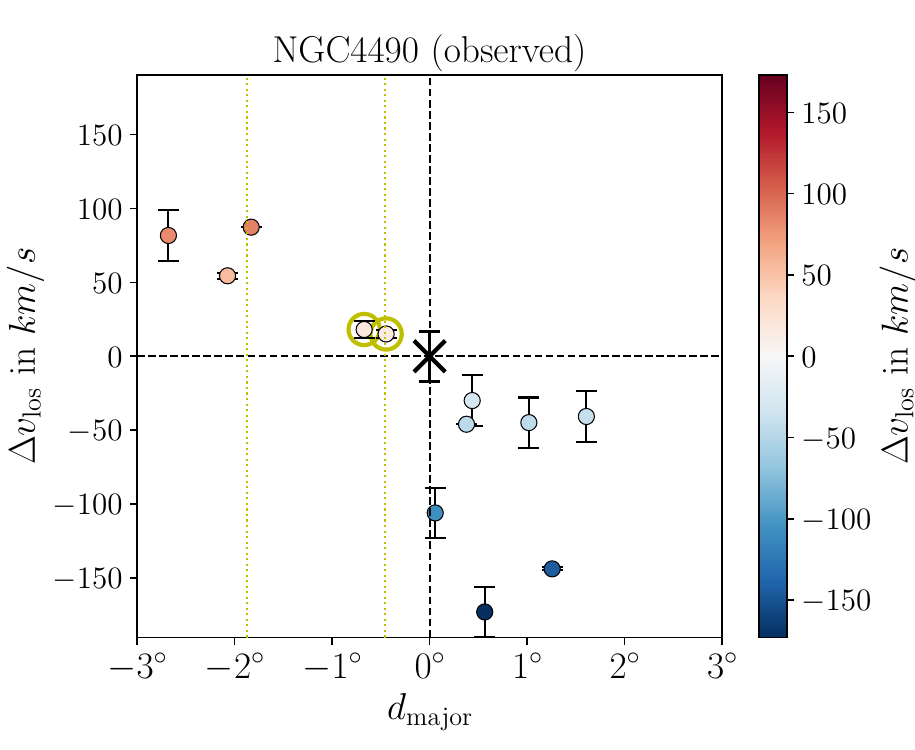}
      \caption{Position-velocity diagram of the NGC\,4490 system. Velocities $\Delta v_\mathrm{los}$\ are given relative to that of NGC\,4490, and positions are in angular separation along the major axis of the system's distribution $d_\mathrm{major}$\ (black dashed line in Fig. \ref{FigOnSky}). The positions of the two objects without available velocities are indicated with vertical yellow dashed lines. Only two quadrants are occupied by the data, consistent with a rotating satellite structure. The likely pair of dwarf galaxies is highlighted by yellow circles.
              }
         \label{FigPV}
   \end{figure}

The barred spiral galaxy NGC\,4490 and its very nearby companion NGC\,4485 are an interacting pair of low-mass galaxies. However, the secondary galaxy, NGC\,4485, has only about $5\%$\ of the $K$-band luminosity of the major galaxy. The pair can thus be considered as a hierarchically merging system, in contrast to a major merger that would require a more equal mass ratio. The galaxies are situated at a distance of 8 to 10\,Mpc \citep{2013AJ....145..101K}, and have a mutual projected separation of less than 10\,kpc. In addition to their disturbed optical morphology indicative of the ongoing interaction, \citet{2023MNRAS.523.3905L} have recently discovered a massive tidal tail in \hi\ that extends along the north-south direction around NGC\,4490 and NGC\,4485, expanding beyond the system's previously known \hi\ envelope.
\citet{2018MNRAS.480.3069P} have investigated the possible dynamics and history of the interacting system via numerical simulations tailored to reproduce the observed system. They find that a 1:8 mass ratio best reproduces the observed morphology and present-day kinematics of the system. However, they also comment on their models having quite low absolute masses, both for the adopted dark matter halo mass and for the observationally better constrained baryonic mass, where the model is 2.6 times less massive than the observed system.

In Fig. \ref{FigOnSky} we show the on-sky distribution of the NGC\,4490 system and the objects associated to it as satellites. We have obtained this list from the original work of \citet{2024MNRAS.528.2805K}, and supplemented it if needed with additional data (such as on the velocity errors) from the Nearby Galaxy Catalog of \citet{2013AJ....145..101K}. Most of these objects are star-forming late-type galaxies. It is apparent that the spatial distribution as projected on the sky is highly flattened. The system's major axis (black dashed line in Fig. \ref{FigOnSky}) is closely aligned with the north-south direction in these Equatorial coordinates. We measured a minor-to-major axis root-mean-square (rms) flattening of the projected distribution of $b/a = 0.381$. 

The position-velocity diagram in Figure \ref{FigPV} shows that the kinematics of the associated objects are strongly correlated along the direction of the major axis of their on-sky positions. All objects with available velocities in the north of NGC\,4490 are approaching us relative to the host (more blue-shifted), while those in the south are receding from us faster than the host (more red-shifted). We measure the degree of kinematic correlation as the number of objects following a coherent velocity trend, such that in this case, we have $N_\mathrm{corr} \geq 12$\ out of 14 objects. 

Since two of the objects associated with NGC\,4490 -- Dw1224+39 and Dw1229+41 -- do not have measured velocities, this degree of kinematic correlation is only a lower limit. Both of these objects are in the south of NGC\,4490, thus a higher value of the kinematic correlation would be obtained if either ($N_\mathrm{corr} = 13$) or both ($N_\mathrm{corr} = 14$) of these have higher recessional velocities (are more red-shifted) than NGC\,4490 (see Fig. \ref{FigPV}).

Assuming an equal probability for positive and negative velocity for each object, the chance to have at least 12 out of 14 objects follow the same velocity trend is 1.29\% for $N_\mathrm{corr} \geq 12$\ out of 14. This drops to 0.18\% and 0.012\% for $N_\mathrm{corr} \geq 13$\ and $N_\mathrm{corr} \geq 14$\ out of 14 objects, respectively. The observed correlation is thus significant and indicative of a physical feature rather than a mere chance occurrence.

Furthermore, the two objects lacking velocities could still turn out to be non-members of the system, for example, if their velocities were measured to exceed the plausible range for the system. In that case, the overall flattening of the system would reduce to $b/a = 0.256$. The change is almost exclusively driven by Dw1224+39, which is a clear spatial outlier, much more off the system's major axis than any other of the objects associated with NGC\,4490.
In this regard it is worth noting that Dw1224+39 is situated only $0.25^\circ$\ from the background starburst galaxy NGC\,4369 that has a heliocentric recession velocity of 1015\,km\,s$^{-1}$, with a corresponding Hubble flow distance placing it about 6\,Mpc behind the here adopted distance to NGC\,4490.

The nature of the newly discovered Plume is less certain due to its large spatial extent, elongation, and assigned velocity from associating it with the surrounding \hi\ \citep{2024MNRAS.528.2805K}. It could be an ultra-diffuse galaxy or a dwarf galaxy that is undergoing tidal disruption. To be conservative, we repeated our analysis here, as well as in comparisons to simulations, after also excluding the Plume. In that case, the overall flattening of the observed system is $b/a = 0.382$\ and its kinematic correlation $N_\mathrm{corr} \geq 11$\ out of 13. In this case the probability to arrive at the observed degree of kinematic correlation by chance is 2.25\% for  $N_\mathrm{corr} \geq 11$\ out of 13 (and 0.34\% or 0.024\% for  $N_\mathrm{corr} \geq 12$\ and  $N_\mathrm{corr} \geq 13$\ out of 13, respectively).

\subsection{A pair of dwarf galaxies}
\label{SectPair}

As can be seen in Figures \ref{FigOnSky} and \ref{FigPV}, two of the galaxies associated with NGC\,4490 are both spatially close to each other and share essentially the same velocity. These are Dw1234+41 and UGC\,7751. They are separated in projection by $0.24^\circ$, which corresponds to a projected physical separation of $d_\mathrm{proj} = 37$\,kpc at the distance of NGC\,4490. Their velocity offset is only $\Delta v_\mathrm{los} = 3\,\mathrm{km\,s}^{-1}$, which is well within their velocity errors of $\delta v =$3 and $6\,\mathrm{km\,s}^{-1}$, respectively. While their distance measurements differ ($D = 8.45\ \mathrm{and}\ 7.90\,$Mpc, respectively) this separation is well within the expected uncertainty on the measurements, which are based on either the Numerical Action Method (for Dw1234+41, see \citealt{2017ApJ...850..207S}) or the Tully-Fisher-Relation method (for UGC\,7751, see \citealt{2013AJ....145..101K}).

We applied the metric defined by \citet{2010ApJ...711..361G} to test whether the two dwarf galaxies could be currently bound to each other and infer the mass required for them in this case. For two point masses to be gravitationally bound, their kinetic energy has to be lower than their total gravitational potential energy. One can define a criterion $b \equiv 2GM_\mathrm{pair}/\Delta r\Delta v^2$, with the physical separation between the objects $\Delta r$, their total velocity difference $\Delta v$, the total mass of the pair $M_\mathrm{pair}$, and the gravitational constant $G$. The system can be considered bound if $b > 1$.

For the present case, we can only reliably consider projected positions and line-of-sight velocities, and do not have an independent estimate of the total mass of the two objects. We thus instead calculated the mass $M_\mathrm{pair}$\ required to arrive at $b > 1$, adopting the projected separation and line-of-sight velocity difference for $\Delta r = d_\mathrm{proj} = 37$\,kpc and $\Delta v = \Delta v_\mathrm{los}$, respectively. The latter two quantities are absolute lower limits, so the calculated mass indicates the minimum mass the pair needs to have to potentially constitute a bound pair\footnote{For the geometrically most-likely case, both quantities would need to be multiplied by $\sqrt{2}$\ for $\Delta r = \sqrt{2} d_\mathrm{proj}$\ and $\Delta v = \sqrt{2} \Delta v_\mathrm{los}$, so the most-likely required mass of the pair for $b = 1$ is $M_\mathrm{pair} = 1.1 \times 10^{8}\,M_\odot$. This is about twice the total baryonic mass, but well below the expected total mass including dark matter.}. For $b = 1$, we obtained $M_\mathrm{pair} = 3.9 \times 10^{7}\,M_\odot$. The $K$-band luminosities of Dw1234+41 and UGC\,7751 are $L_K = 1.8 \times 10^7\,L_\odot$\ and $3.1 \times 10^7\,L_\odot$, respectively, and UGC\,7751 has additionally a reported \hi\ mass of $M_\mathrm{\hi} = 3.8 \times 10^7\,M_\odot$ \citep{2013AJ....145..101K}. Assuming $M/L_K \approx 0.5$, the system's baryonic mass alone is thus already sufficient to make it potentially bound. If the dwarf galaxies are embedded in dark halos, one can expect these to be about two orders of magnitude more massive than the baryonic component for dwarf galaxies of this scale. Thus, with $M_\mathrm{pair} \geq 4 \times 10^9\,M_\odot$\ their full 3D velocity difference could still be ten times higher ($\sim 30\,\mathrm{km\,s}^{-1}$), or they could have a substantial separation along the line of sight (or some combination thereof) while still very plausibly being a bound pair of dwarf galaxies.

\subsection{The larger-scale context of NGC\,4490}

   \begin{figure*}
   \centering
   \includegraphics[width=0.44\hsize]{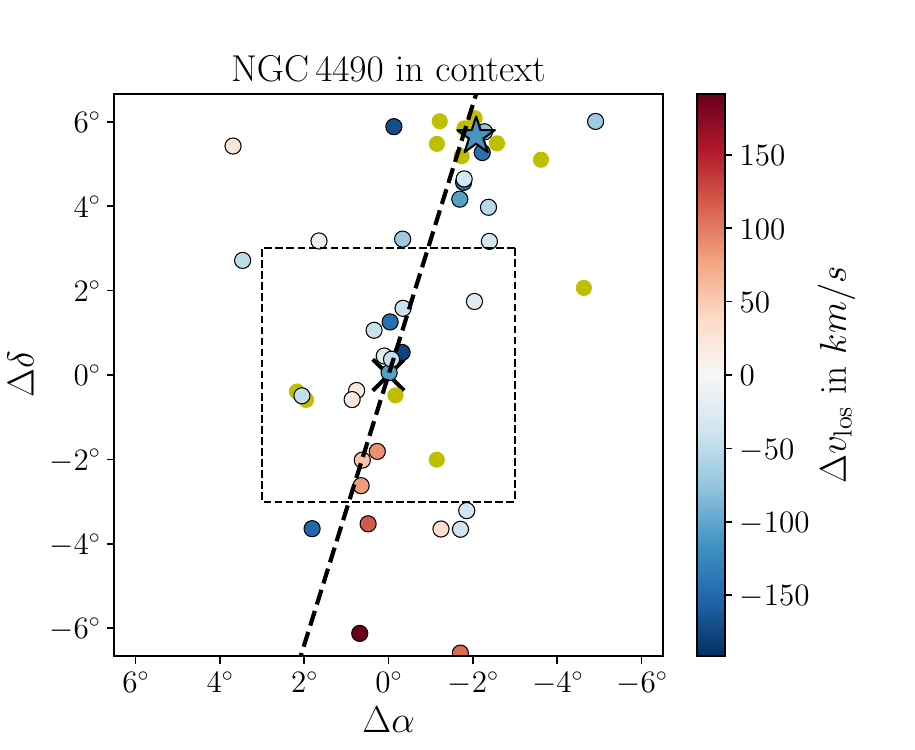}
   \includegraphics[width=0.275\hsize]{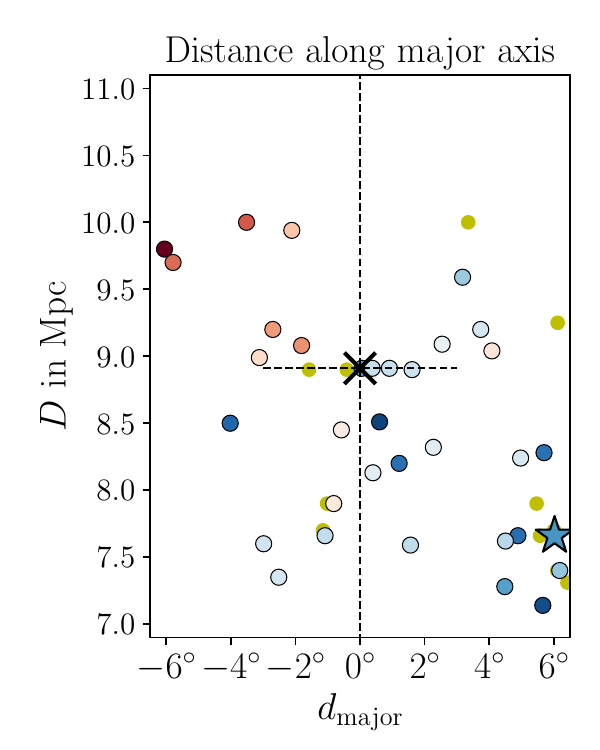}
   \includegraphics[width=0.275\hsize]{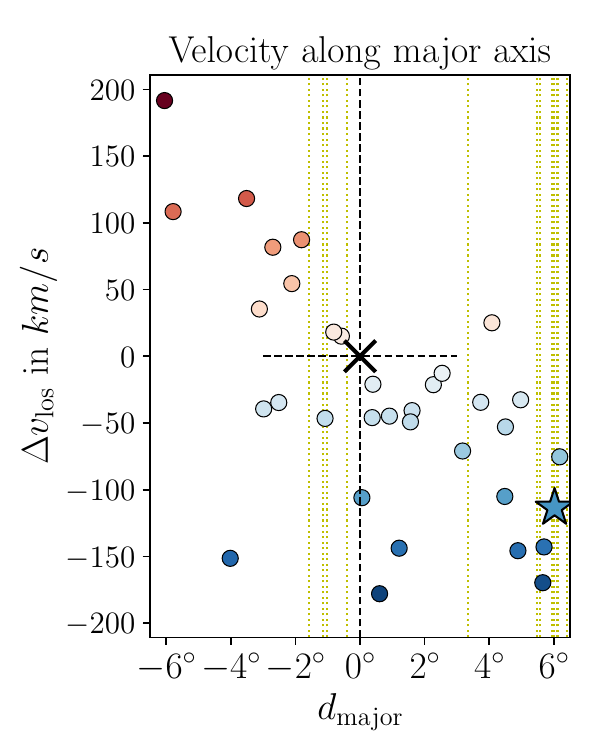}
      \caption{Larger-scale context of the region surrounding NGC\,4490 (left panel). Galaxies are color-coded by their velocity relative to that of NGC\,4490, while those without velocity measurements are shown in yellow. The star symbol highlights the nearby, brighter spiral galaxy M106. The dashed black line indicates the major axis of this larger-scale distribution of galaxies. It aligns with the orientation of the spatially flattened distribution of objects close to NGC\,4490 of Fig. \ref{FigOnSky}. The dashed square indicates the dimensions of that figure.
      The middle panel shows the distance of galaxies along this major axis, while the right panel shows the velocity of galaxies along this major axis. 
      The dashed horizontal lines indicate the extent of the dashed square in the left panel.
              }
         \label{FigContext}
   \end{figure*}

Fig. \ref{FigContext} shows the larger-scale context of the region surrounding NGC\,4490. It includes galaxies selected from the compilation of  \citet{2013AJ....145..101K} to be within the distance range of 7 to 11\,Mpc and having velocities relative to that of NGC\,4490 of $\pm 300\,\mathrm{km\,s}^{-1}$.
It is apparent that the spatially flattened distribution of objects aligns well with the larger-scale distribution of surrounding galaxies. In particular, the more luminous spiral galaxy M106 (NGC\,4258) with its own entourage of satellites resides towards the north of NGC\,4490 at a separation of $6^\circ$. At the distance of NGC\,4490, this corresponds to about 0.9\,Mpc. In distance, M106 is closer to us by about 1.25\,Mpc and has a lower recessional velocity. The latter agrees with the velocity trend of the objects in the direct vicinity of NGC\,4490. In fact, the velocity coherence of the systems close to NGC\,4490 appears to continue at larger distances. This could be indicative of a connection to the larger-scale cosmic web, with NGC\,4490 and M106 being part of the same cosmic filament that is oriented perpendicular and somewhat inclined relative to our line-of-sight. With galaxies potentially moving along such a filament, this can result in the accretion of galaxies and groups of galaxies \citep{2024arXiv240416110J} from preferential directions \citep{2011MNRAS.411.1525L,2019MNRAS.490.3786L}, but could also explain the orientation of the interaction between NGC\,4490 and NGC\,4485 that could in turn form a correlated satellite system \citep{2011A&A...532A.118P}.

Several effects related to the larger-scale motion and distribution of galaxies could induce a velocity coherence. One possibility is that the observed gradient in line-of-sight velocities (Fig. \ref{FigPV}) is caused by a transverse motion of the system \citep[an effect explored already by][to measure the proper motion of the Large Magellanic Cloud]{1920MNRAS..80..782H}. Such a transverse motion with velocity $v_\mathrm{trans}$\ would increasingly contribute to the line-of-sight component of the velocity $\Delta v_\mathrm{los}(d_\mathrm{major})$\ for larger angular separations $d_\mathrm{major}$\ from the center of the system. Here we assumed this motion would align with the major axis, because that direction maximizes the velocity coherence. A linear least-squares-fit to the data in Fig. \ref{FigPV} returned a velocity gradient of $48\,\mathrm{km\,s}^{-1}\,\mathrm{deg}^{-1}$. Over the three degrees in radius considered around NGC\,4490, this would imply a transverse velocity of $v_\mathrm{trans} = \frac{144\,\mathrm{km\,s}^{-1}}{\sin(3^\circ)} = 2750\,\mathrm{km\,s}^{-1}$. This is unrealistically high for a peculiar motion \citep[e.g.,][]{2024MNRAS.527.3788H}, and thus we dismiss the possible that the strong observed velocity coherence is caused by such transverse motion.

Alternatively, the coherence in velocities could be indicative of motion along cosmic filaments. It can generally be expected that dwarf galaxies tend to stream towards a host along such filaments. If this process happens along two filaments from opposite sides of the host -- which appears plausible in this case given the larger-scale distribution of galaxies shown in Fig. \ref{FigContext} -- this could induce a velocity coherence: a filament approaching NGC\,4490 from behind (i.e., from larger distances) would be moving towards us relative to NGC\,4490 and thus be less redshifted, while a filament in front of the host would be moving towards the host and thus be more redshifted than the host. For the orientation of the filament around NGC\,4490, we adopted the line connecting NGC\,4490 and M106. M106 is 1.25\,Mpc closer than NGC\,4490 and offset by $6^\circ$\ towards the north (0.9\,Mpc at the distance of NGC\,4490). Thus from north to south in Figures \ref{FigOnSky} and \ref{FigContext}, the filament would extend from the closer M106 to the more distant NGC\,4490 to even larger distances, with a distance gradient of about 200\,kpc per degree. Yet, relative to NGC\,4490, the velocities of the dwarfs in the north are approaching us and those in the south are receding from us. This is the opposite of the expected trend, and would mean that the dwarfs would not be accreting onto NGC\,4490 along the filament, but streaming away from it on both sides. This is at odds with finding the dwarfs spatially clustered around NGC\,4490 at this time.

Finally, another possibility is that the velocity coherence is governed by the Hubble flow. In this case, the orientation of the filament is consistent. With M106 being closer to us by 1.25\,Mpc, that direction would be less redshifted, in line with the observed trend of velocities. However, adopting $H_0 = 70\,\mathrm{km\,s}^{-1}\,\mathrm{Mpc}^{-1}$, a distance gradient of 200\,kpc per degree corresponds to a Hubble-expansion-induced velocity gradient of only $14\,\mathrm{km\,s}^{-1}$\ per degree, and could thus only account for a fraction of the observed velocity gradient of $48\,\mathrm{km\,s}^{-1}\,\mathrm{deg}^{-1}$. In other words, the velocity range of the position-velocity diagram in Fig. \ref{FigPV} of about 150\,km\,s$^{-1}$\ over the three-degree field in radius would require a distance difference of $\pm 2$\,Mpc if it were merely due to the Hubble flow, which is in conflict with the distance measurements to the considered objects.

While each of these effects can contribute in one way or another to the observed velocities, we consider it implausible that they are the cause for the strong observed velocity coherence in the immediate vicinity of NGC\,4490. However, we do caution that improved observational constraints, and a more complete census of the galaxy system associated with M106, would be highly desirable to develop a better understanding of the larger-scale structure and dynamics of the region.


\section{Analogs in cosmological simulations}

   \begin{figure}
   \centering
   \includegraphics[width=0.8\hsize]{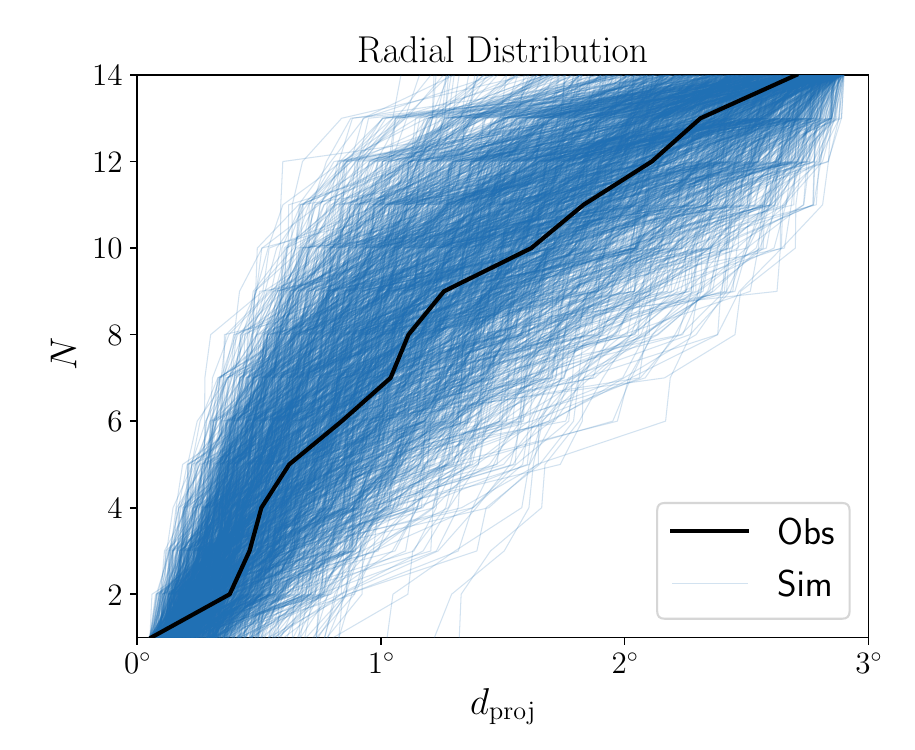}
   \includegraphics[width=0.8\hsize]{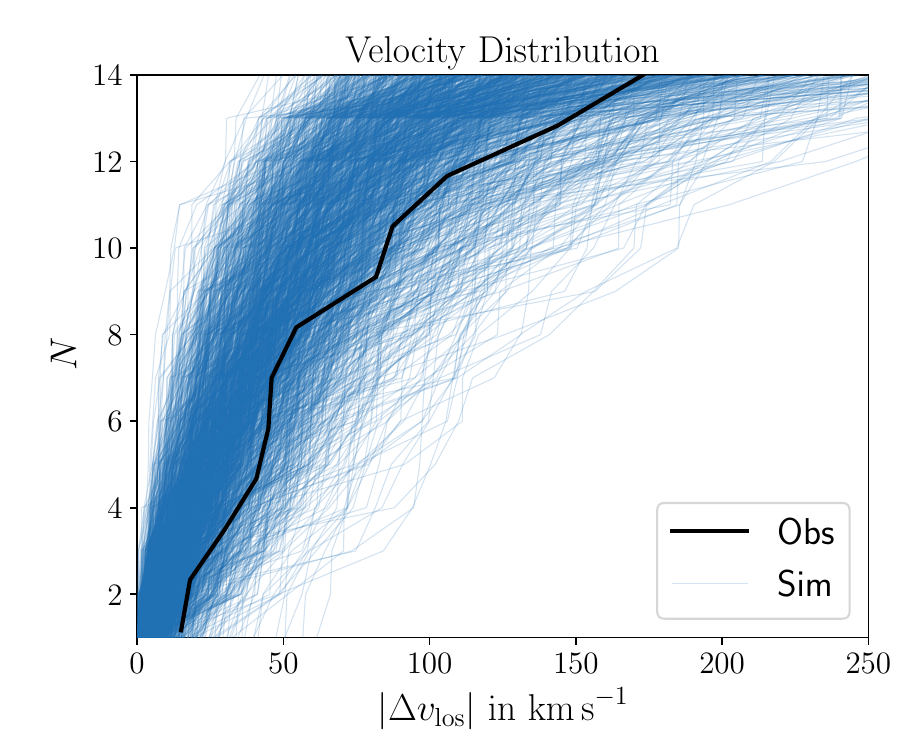}
   \includegraphics[width=0.8\hsize]{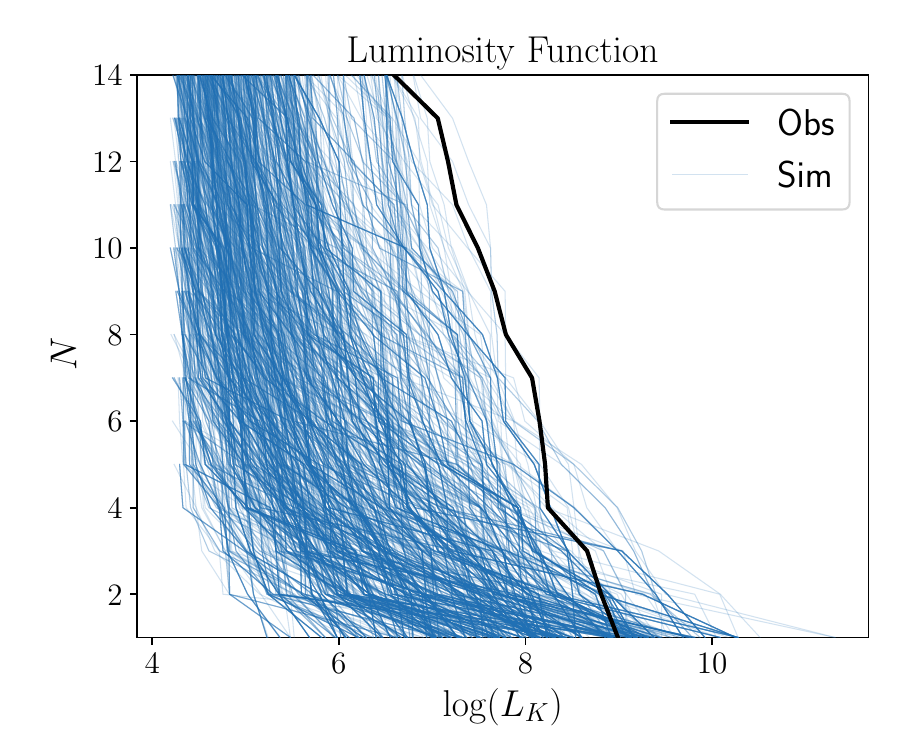}
   \caption{Radial distribution of objects in simulated analogs (top panel), distribution in line-of-sight velocities (middle panel), and their luminosity distribution in $\log_{10}\left(L_K\right)$\ (bottom panel), compared to that of the observed NGC\,4490 system (black lines). The selected simulation analogs closely follow the observed system's spatial distribution and are consistent with its velocity distribution. However, they are on average less luminous and only barely approach the observed luminosity function within their scatter, potentially indicating a too-many-satellites problem for NGC\,4490. 
   }
              \label{FigSimProperties}
    \end{figure}

For our comparison, we used the IllustrisTNG project \citep{2018MNRAS.475..676S, 2018MNRAS.475..648P}. It is a suite of detailed hydrodynamical cosmological simulations that adopts cosmological parameters consistent with those reported by \citet{2016A&A...594A..13P}. Specifically, we searched for analogs of the NGC\,4490 system in the hydrodynamical TNG50-1 run \citep{2019MNRAS.490.3234N}, using the publicly available galaxy catalog \citep{2019ComAC...6....2N} at redshift zero. The box size of this run is $35\,\mathrm{Mpc\,}{h}^{-1}$ and the dark matter particle mass is $3.1 \times 10^{5}\,\mathrm{M} _\odot\,h^{-1}$. The latter provides the highest resolution of all runs, a requirement due to the rather low-mass host we study. Nevertheless, by utilizing this IllustrisTNG run, which employs the same physics implementation as the larger-volume runs, we ensured maximal comparability with previous comparisons with the satellite planes the Milky Way \citep{2020MNRAS.491.3042P}, M31 \citep{2021ApJ...923...42P}, and Centaurus\,A \citep{2021A&A...645L...5M} that were based on TNG100-1. 

The virial mass of the NGC\,4490 group is only poorly constrained. \cite{2024MNRAS.528.2805K} report $M_\mathrm{total} = (1.37 \pm 0.43) \times 10^{12}\,M_\odot$\ based on estimating the total Newtonian mass via the velocity offset between each individual assumed satellite and the host. In contrast, the best-matching dynamical model of \citet{2018MNRAS.480.3069P} has a total halo mass of only $M_\mathrm{NGC\,4490, halo} = 1.6 \times 10^{10}\,M_\odot$ for NGC\,4490 itself. However, these authors comment on the latter being quite low, and that their model even has 2.6 times less total baryonic mass than the observed system. Via abundance matching, they obtain an alternative mass of $M_\mathrm{NGC\,4490, halo} = 2.6 \times 10^{11}\,M_\odot$.

Thus in the simulation, we selected all potential hosts from a rather broad range in virial mass of $2 \times 10^{11}\,M_\odot < M_\mathrm{vir} < 2 \times 10^{12}\,M_\odot$. These were required to be isolated by rejecting all potential hosts that have a neighboring halo of virial mass $\geq 2 \times 10^{11}\,M_\odot$\ within 1\,Mpc. This is motivated by the closest nearby massive neighbor, M106, at a separation of approximately 1.5\,Mpc.

All subhalos within 1.5\,Mpc of the host were pre-selected. They were ranked by absolute $K$-band magnitude $M_\mathrm{K}$\ or, in the case of dark subhalos, by their total dark matter mass (to account for faint satellites that do not have a resolved stellar component in the simulation). Hosts that have a nearby satellite with a stellar mass exceeding 25\% of that of the host were also rejected to avoid major-merger-like cases -- while NGC\,4490 does interact with NGC\,4485, the latter has only about 5\% of the K-band luminosity of the former. 
For each host, ten different random view directions were drawn \footnote{To ensure that our obtained frequencies are not dominated by the choice of ten random view directions, we repeated our analysis also with sampling 100 random views per host. The obtained frequencies are consistent with the ones for ten view directions that are reported in the following (e.g., $f_\mathrm{flat} = 7.14\%$, $f_\mathrm{corr} = 3.12\%$, $f_\mathrm{both} = 0.24\%$). However, since some of these 100 directions will be quite similar and thus not independent, we restricted our comparisons to ten views per system.}. For each, the host and its satellites were placed at the adopted distance of the observed NGC\,4490 group of $D =  8.91\,\mathrm{Mpc}$\ from the observer along this direction. Satellites were then selected within a circular aperture of $2.9^\circ$\ around the host, which corresponds to 450\,kpc at this distance and is motivated by the projected separation of the furthest object considered to be associated with NGC\,4490 by \cite{2024MNRAS.528.2805K}, UGC\,7719 at a projected separation of 420\,kpc. 
The $N$\ top-ranked satellites within this field were selected and their projected positions and line-of-sight velocities relative to the mock observer were recorded. This data was then treated analogously to the observed NGC\,4490 system by measuring the projected rms minor-to-major axis flattening $b/a$, and counting the number $N_\mathrm{corr}$\ of objects that are kinematically correlated relative to the determined major axis of the system.

For our fiducial comparison, we selected $N = 14$\ satellites. As shown in Figure \ref{FigSimProperties}, the selected systems are consistent in their radial distribution with the observed one. Their mean line-of-sight velocity dispersion of $\sigma_v = 65  \pm 23\,\mathrm{km\,s}^{-1}$\ is also broadly consistent with, albeit somewhat lower than, that of the observed objects ($\sigma_v = 85\,\mathrm{km\,s}^{-1}$). This is to some degree expected as the simulated systems do not contain measurement errors (of up to $\delta v_\mathrm{los} = 17\,\mathrm{km\,s}^{-1}$), but can also indicate that the observed system does indeed reside in a more massive dark matter halo, or might be contaminated by fore- or background objects. We note, however, that the simulation selection did include objects up to 1.5\,Mpc in front of or behind the host, which exceeds the maximum difference in distance $D$\ between the observed host NGC\,4490 and any of its associated objects by 50\%.

The simulated objects tend to be substantially less luminous than those in the observed NGC\,4490 group. This might be related to the ‘too-many-satellites’ issue that was also reported for the dwarf galaxies in the MATLAS (Mass Assembly of early-Type Galaxies with their fine Structures) survey  compared to the TNG simulation 
(Kanehisa et al. in prep.).
Given that the considered dwarf galaxies are at the lower end of the TNG resolution and are strongly dark matter dominated, we assume that the overall phase-space behavior of the subhalos is not affected by this mismatch in luminosity and remains representative of the $\Lambda$CDM model expectations.

\section{Results}

   \begin{figure*}
   \centering
   \includegraphics[width=0.49\hsize]{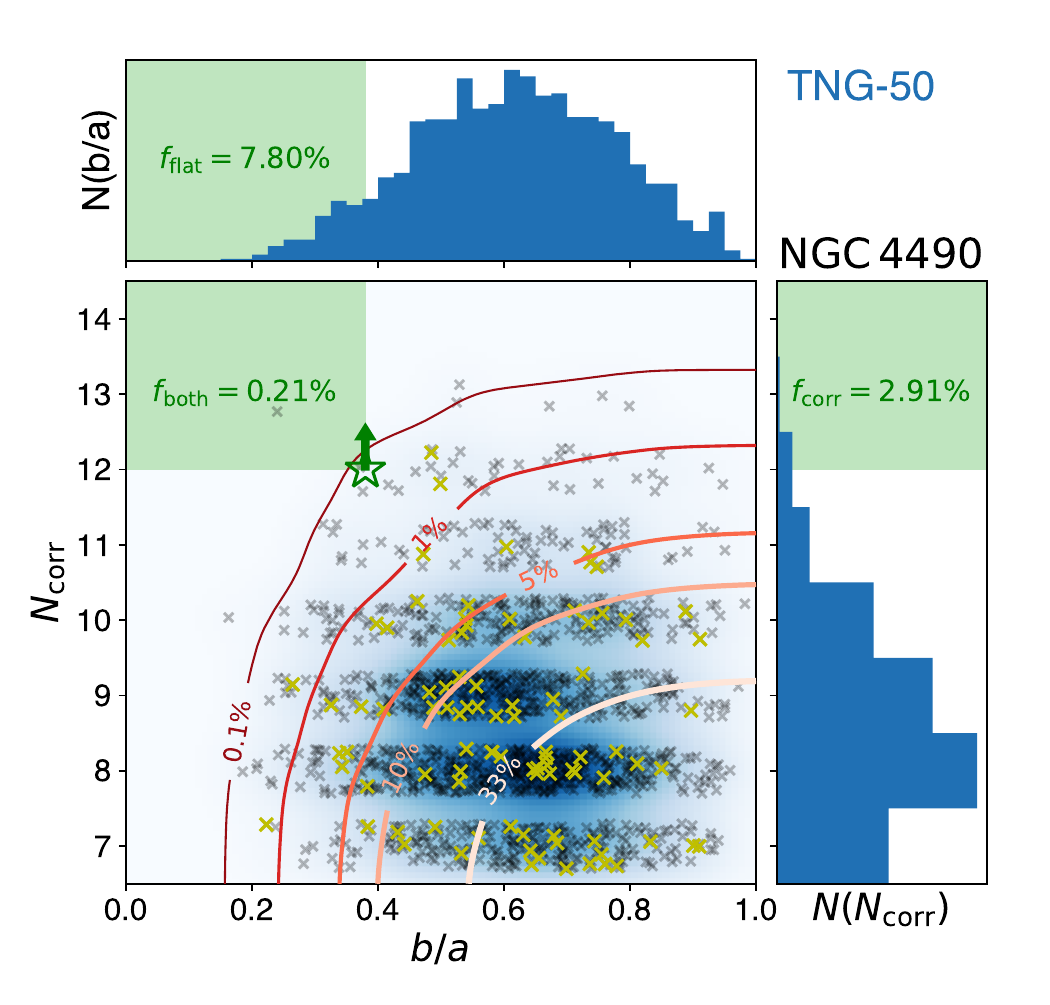}
   \includegraphics[width=0.49\hsize]{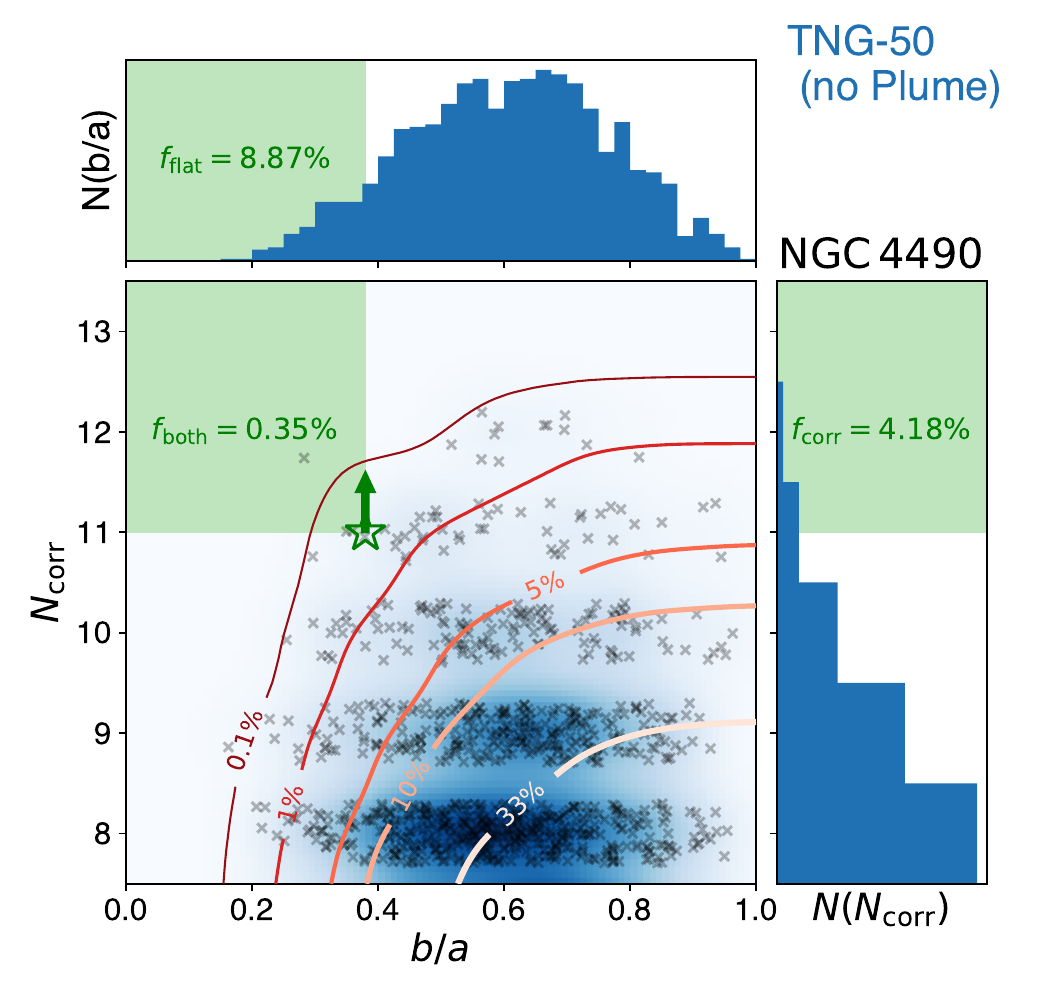}
   \caption{Comparison of on-sky minor-to-major axis flattening $b/a$\ and number of kinematically correlated objects $N_\mathrm{corr}$\ for systems considering 14 satellites in the IllustrisTNG-50 simulation (left panel) and 13 satellites to consider the case without the newly discovered Plume (right panel), with the observed values for NGC\,4490 (green point). Due to the lack of kinematic information for two of the observed objects, $N_\mathrm{corr} = 12$\ out of 14 is only a lower limit, as indicated by the green arrow. Each simulated system is shown as a black cross (with small random offsets in the $N_\mathrm{corr}$\ to better illustrate the distribution despite this being a natural number),  and their underlying distribution as a blue color map. Red contours indicate the fraction of systems more flattened and more kinematically correlated, in levels of 0.1, 1, 5, 10, and 33\% of the simulated analogs. The fraction $f_\mathrm{both}$\ of systems that are both as flattened ($b/a$ < 0.38) and as kinematically correlated ($N_\mathrm{corr} \geq 12$) are given in green.
    The distributions in $b/a$\ and $N_\mathrm{corr}$\ individually and fractions of analogs as extreme as the observed system in these quantities are also shown as histograms along the axes. 
    Systems with an NGC\,4485 analog close to the host are highlighted in yellow in the left panel.
   }
              \label{FigSimFrequencies}
    \end{figure*}

For our fiducial host and satellite selections, we obtained a total of 141 hosts, each observed from ten random directions. The distribution on flattening and kinematic correlation of these systems is shown in Fig. \ref{FigSimFrequencies}.  The expected flattening for the $\Lambda$CDM analogs peaks around $b/a = 0.6$, which is considerably wider than the observed value of 0.38. The typical kinematic correlation of these simulated systems is $N_\mathrm{corr} = 8$\ out of 14 satellites, as expected for largely uncorrelated kinematics.

In contrast, the observed system is both more flattened and more correlated. Only $f_\mathrm{flat} = 7.8\%$\ of the simulated analogs have $b/a < 0.38$, and only $f_\mathrm{corr} = 2.9\%$\ have $N_\mathrm{corr} \geq 12$. If either one of the two observed objects associated with NGC\,4490 without current velocity measurements turns out to contribute to the kinematic coherence, the latter frequency would drop to $f_\mathrm{corr} = 0.4\%$\ for $N_\mathrm{corr} \geq 13$. In none of the simulated analogs, all of the objects partake in the same kinematic trend ($N_\mathrm{corr} = 14$, $f_\mathrm{corr} = 0.0\%$). The simulated systems thus are about twice as likely to show the observed degree of kinematic correlation than completely randomly drawn velocities, indicative of the degree of overall phase-space correlation present in $\Lambda$CDM\ satellite systems.

To reproduce the observed NGC\,4490 system, both the observed flattening and the kinematic coherence should be jointly realized in a simulated analog. This occurs in only $f_\mathrm{both} =  0.21\%$\ of the cases for $N_\mathrm{corr} \geq 12$, and would be reduced by a factor of three if one of the two objects without velocities turns out to follow the kinematic trend.

Since the nature of the newly discovered Plume is debatable, we repeated our analysis without it and thus compare to only 13 simulated satellites per host (see right panel of Fig. \ref{FigSimFrequencies}). In this case, the overall flattening of the observed system is reproduced in $f_\mathrm{flat} = 8.9\%$\ of the simulated systems, and the kinematic coherence of $N_\mathrm{corr} = 11$\ in $f_\mathrm{corr} = 4.2\%$ (again about twice the value for randomly drawn velocities). The value of $f_\mathrm{corr}$\ drops to $0.9\%$\ for $N_\mathrm{corr} = 12$. Only $f_\mathrm{both} = 0.35\%$\ of the simulated systems reproduce both the observed flattening and kinematic correlation simultaneously ($f_\mathrm{both} = 0.07\%$\ for $N_\mathrm{corr} = 12$). The rarity of analogs of the proposed satellite plane around NGC\,4490 is thus largely robust to in- or exclusion of this particular object.

Repeating the analysis with the additional requirement that all selected satellites must contain a luminous stellar component, we obtained 1056 analogs with a sufficient number of luminous satellites. However, the previously obtained frequencies do not change significantly. We find $f_\mathrm{flat} = 7.9\%$\ of the simulated analogs are at least as flattened as observed, and $f_\mathrm{corr} = 2.8\%$\ are at least as kinematically correlated ($N_\mathrm{corr} \geq 12$). Simultaneously reproducing both requirements happens in $f_\mathrm{both} =  0.28\%$\ of the systems.

\subsection{Effect of host interacting with a nearby galaxy}

NGC\,4490 is currently interacting with the nearby galaxy NGC\,4485. This warrants the question whether such a configuration might affect the frequency or the properties of satellite planes in the associated satellite systems in the simulation. To investigate this possibility, we identified those simulated analogs which contain a galaxy similarly luminous as NGC\,4485 ($8.0 < \log(L_K) < 10.0$) within an angle of $0.3^\circ$\ and a three-dimensional separation of less than 50\,kpc from the host. 

The cases that  fulfill these consitions are highlighted in the left panel of Fig. \ref{FigSimFrequencies}. They do not show any distinct difference compared to the overall distribution, in neither flattening nor kinematic correlation. None of these analogs containing a NGC\,4485-like nearby object are both as flattened and as kinematically correlated as the observed system. This apparent absence of an effect on the prevalence of satellite planes is in line with the results by \citet{2023MNRAS.524..952K}, who found that major mergers are not conducive to the formation of correlated satellite planes.

\subsection{Best simulated analog}

   \begin{figure}
   \centering
   \includegraphics[width=\hsize]{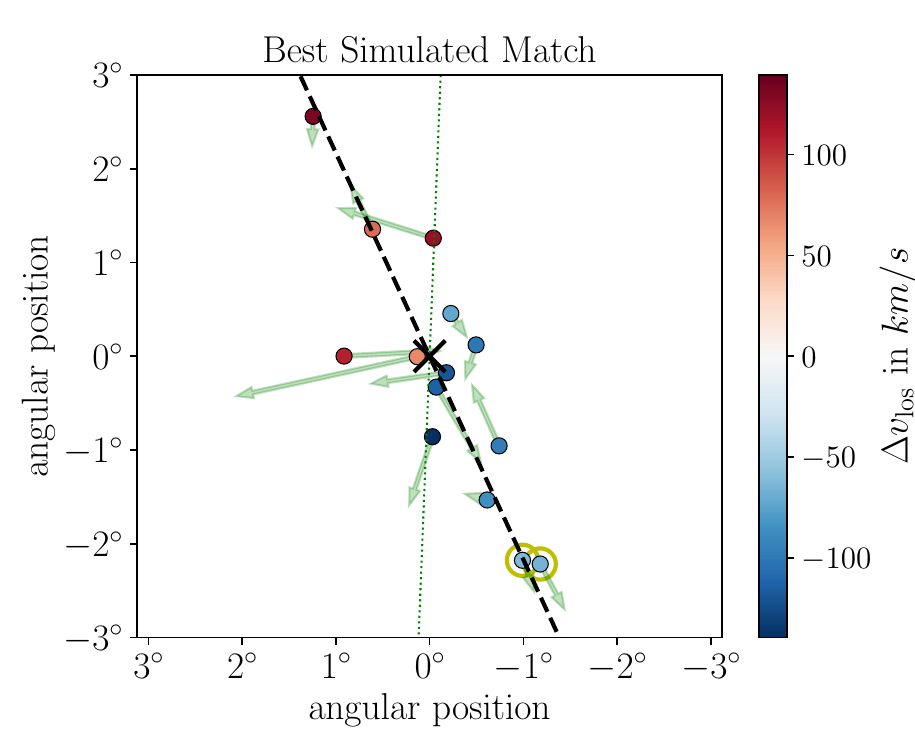}
      \caption{On-sky distribution of objects for the best-matching simulated analog of the observed system. The dashed line indicates the major axis of the projected distribution, the axis ratio is $b/a = 0.24$. Out of the 14 considered galaxies, 13 show a coherent velocity trend, and this would increase to 14 if the system were split along the dashed green line. The close pair of dwarf galaxies is highlighted with yellow circles. They are moving in the same direction, as indicated by the green arrows showing the (observationally inaccessible) on-sky motion of the simulated galaxies.
              }
         \label{FigBestAnalog}
   \end{figure}

Only three simulated analogs out of 141 match the observed system's flattening and correlation, and only for one of their ten random view directions each. The best of these analogs (simultaneously most flattened and most kinematically correlated) is shown in Fig. \ref{FigBestAnalog}.

The host of this system has a $K$-band magnitude of $M_\mathrm{K} -24.2$, a virial mass of $M_\mathrm{vir} = 1.39 \times 10^{12}\,M_\odot$, and a stellar mass of $M_\mathrm{*} = 5.36 \times 10^{10}\,M_\odot$. This makes it somewhat more luminous than the observed NGC\,4490 galaxy, but places it exactly at the total Newtonian mass estimated by \cite{2024MNRAS.528.2805K}. The associated objects are, in this view direction, strongly flattened with $b/a = 0.24$, and $N_\mathrm{corr} = 13$\ out of 14 follow a coherent velocity trend. The green line in Fig. \ref{FigBestAnalog} indicates how the positions could be split to even arrive at a $N_\mathrm{corr} = 14$, though this split is not aligned with the system's major axis anymore. 

Particularly intriguingly in light of the discussion in Sect. \ref{SectPair} is that the two objects at the bottom of the figure with negative velocities of order $60\,\mathrm{km\,s}^{-1}$\ appear to be a potential galaxy pair. In fact, they share strikingly similar properties to the Dw1234+41 and UGC\,7751 pair. They have similar magnitudes of  $M_K = -14.0$\ and $-13.6$\ ($L_K = 0.94 \times 10^7\,M_\odot$\ and $0.64 \times 10^7\,M_\odot$), are separated by only 36.4 kpc in 3D, and have a total velocity difference of only $34.5\,\mathrm{km\,s}^{-1}$.  Their total mass is $M_\mathrm{pair} = 3.4 \times 10^9\,M_\odot$. Formally, this gives a criterion of $b = 0.675$, which is just barely below the threshold of unity for a bound pair, though can be considered close enough given the inaccuracies in the estimation method employed such as the assumption of point masses.


\section{Discussion and conclusions}

In this work, we have investigated the recently discovered plane-of-satellites around the barred spiral galaxy NGC\,4490 in light of $\Lambda$CDM cosmology. For that purpose, we have measured the two-dimensional flattening of the observed satellite system, as well as its kinematical coherence. We find that the minor-to-major axis ratio is $b/a = 0.38$\ based on all 14 possible satellites, but goes down to $b/a = 0.26$\ if we only consider the 12 confirmed satellite galaxies. Out of these confirmed satellites, all seem to behave as expected from a co-rotating plane-of-satellite seen edge-on, namely that they are redshifted on the one side and blueshifted on the other side with respect to the host galaxy. Such a velocity trend occurs in only 1.3\% of the cases by chance (i.e., finding $N_{corr}\geq12$ out of 14). This estimate must be considered a lower limit, because the two dwarf galaxy candidates lacking velocity measurements may follow the velocity trend as well. In this context, it is interesting to note that one of these dwarf galaxy candidates is a clear spatial outlier and is the object furthest away from the major axis of the system.

Of the observed satellites, two seem to form a pair -- Dw1234+41 and NGC\,7751 --, with a close on-sky separation (37\,kpc) and a velocity difference of only 3 km s$^{-1}$. We tested whether this pair is bound using the metric defined by \citet{2010ApJ...711..361G}. However, because we do not have the dark matter mass of these dwarf galaxies, we instead calculated the total mass of the pair required to make them bound. The measured baryonic mass (in the form of stars and gas) is already sufficient to make the two galaxies bound, and their dark matter content will only add to that. We conclude that in all likelihood this is a bound pair of dwarf galaxies.

The larger-scale context of the group might hold clues on the origin of the observed flattening and correlation. The flattening continues at larger distances and is well aligned towards the nearby, more massive neighbor M106 at a distance of 1.5\,Mpc from NGC\,4490. The kinematic trend, too, appears to continue outside of the immediate vicinity of NGC\,4490. This makes it plausible that the observed structure could be related to the surrounding cosmic web, and might be indicative of the dominant direction of motion of the local cosmic environment such as galaxies moving and colliding along a filament.

To address how the observed phase-space correlation of the NGC\,4490 satellite system compares to cosmological $\Lambda$CDM simulations we have used the Illustris TNG-50 suite of hydrodynamical simulations and searched for analogs of the NGC\,4490 group. While the analogs match the observed radial and velocity distributions, they are less luminous than the observed satellite system. This is consistent with previous findings on M\,83 and the MATLAS sample and may indicate a too-many satellite problem.

Taking these simulated satellites we projected the analogs into 2D and applied the same measurements as we did for the observed system, namely measured the flattening and the kinematic coherence. Jointly considering these properties, we found only 0.2\% of the simulated systems host a similar satellite system as observed. And while the best matched system sports a similar coherence (13/14) and flattening ($b/a=0.24$), its three-dimensional motions do not suggest a fully co-rotating structure. Interestingly, though, we find a pair of subhalos similar to the observed pair of dwarf galaxies. And more, their three-dimensional motion is aligned with the major axis of the system -- as expected for a co-rotating plane. Such small dwarf galaxy groups may thus point towards the origin of co-rotating systems \citep{2024arXiv240416110J}.
Neither restricting the analysis to only luminous satellites nor requiring the presence of an analog to NGC\,4485 (i.e., a luminous galaxy close to the host) appeared to be conducive to the presence of a flattened and kinematically correlated overall satellite distribution in the simulations.

The NGC\,4490 group is a 3$\sigma$ outlier based on our analysis employing Illustris TNG-50. This is consistent with estimates found for other galaxies in the Local Volume. We note again that this comparison of the NGC\,4490 system is only a lower limit, because the two remaining dwarf galaxy candidates could increase the kinematic coherence further and with that lower the probability of finding analogs in the simulation with equal or more extreme properties. Namely, it would go to zero if the two were to follow the kinematic trend observed for the other dwarfs or if they don't belong to the system at all. The NGC\,4490 system, together with previous studies of groups within the Local Volume, show that the plane-of-satellite problem remains a severe challenge for $\Lambda$CDM cosmology.

\begin{acknowledgements}
Marcel S. Pawlowski acknowledges funding via a Leibniz-Junior Research Group (project number J94/2020).
Oliver Müller is grateful to the Swiss National Science Foundation for financial support under the grant number PZ00P2\_202104. 
We thank the anonymous referee for their constructive input which has helped us improve the manuscript.
\end{acknowledgements}

%
   \bibliographystyle{aa} 
   \bibliography{NGC4490bib} 
%

\end{document}